\documentclass[twocolumn,showkeys,showpacs,prb]{revtex4}
\usepackage{amsfonts, amsmath, latexsym}
\usepackage[dvips]{graphicx}

\begin{document}

\title{Sampling diffusive transition paths}
\author{Thomas F. Miller III}
\email{tfmiller@berkeley.edu} 
\author{Cristian Predescu}
\email{cpredescu@comcast.net} 

\affiliation{Department of Chemistry and Kenneth S. Pitzer Center for Theoretical Chemistry, University of California, Berkeley, California 94720}

\date{\today}
\begin{abstract}
We address the problem of sampling double-ended diffusive paths.  The ensemble of paths is expressed using a symmetric version of the Onsager-Machlup formula, which only requires evaluation of the force field and which, upon direct time discretization, gives rise to a symmetric integrator that is accurate to second order.  Efficiently sampling this ensemble requires avoiding the well-known stiffness problem associated with sampling infinitesimal Brownian increments of the path, as well as a different type of stiffness associated with sampling the coarse features of long paths.
The fine-feature sampling stiffness is eliminated with the use of the fast sampling algorithm (FSA), and the coarse-feature sampling stiffness is avoided by introducing the sliding and sampling (S\&S) algorithm.
A key feature of the S\&S algorithm is that it enables massively parallel computers to sample diffusive trajectories that are long in time.
We use the algorithm to sample the transition path ensemble for the structural interconversion of the 38-atom Lennard-Jones cluster at low temperature.
\end{abstract}
\maketitle

\section{Introduction}

Transition path sampling (TPS) is a successful approach to characterizing 
rare event processes for which \textit{a priori} knowledge of the location of, or even existence of, 
a single well-defined transition state is not available.  
It addresses the rare event problem by focusing computational resources on the small subset of trajectory space that exhibits dynamically interesting events.  
 The development of the approach into a computationally effective tool is due largely to the efforts of Chandler and 
coworkers,\cite{Del98a, Del98b, Del98c, Del99,TPSrev1,TPSrev2} 
and it is now widely used to study the mechanisms and rates of reactive processes.

The TPS method prescribes Metropolis Monte Carlo sampling of the transition path ensemble, which includes only those dynamical trajectories of a given length in time that begin in a prescribed ``reactant'' region and terminate in prescribed ``product'' region.  A central feature of the TPS method is the preservation of the Boltzmann distribution through detailed balance.  

As in all Metropolis algorithms, TPS involves a proposal step, in which a new dynamical trajectory is constructed, and then an acceptance/rejection step.  New trajectories can be proposed in a variety of ways.  One approach is to employ open-ended trajectory algorithms, which generate new trajectories by directly integrating the dynamics of the system.  Examples of this approach include the widely used ``shooting''  and ``shifting''  techniques.\cite{Del98b}
An alternative approach employs double-ended trajectory algorithms, in which the dynamical trajectories are treated as extended geometrical objects, or chains-of-states,\cite{Pra86,Del98a,Stu04} 
and sampled using path integral Monte Carlo methods.

In practice, two factors have contributed to the predominance of TPS applications that employ opened-ended trajectories.\cite{Del98b,TPSrev1,TPSrev2} 
Firstly, for processes that react by way of a single barrier-crossing event, the open-ended shooting and shifting algorithms are very effective in providing decorrelated members of the transition path ensemble, thus rendering a double-ended approach unnecessary.
 Secondly, the path integral sampling problem that arises in a double-ended formulation of TPS has appeared to be prohibitively difficult.
 
However, strong motivation for the development of double-ended TPS algorithms does exist.
Many complex systems, including many biologically relevant systems, undergo transition processes that involve slow, diffusive barrier crossing dynamics and exhibit the possibility of metastable intermediates. 
These conditions are known to cause
difficulty for the open-ended TPS algorithms,\cite{TPSrev1} but they are naturally overcome in double-ended methods which enable the trajectory endpoint positions to be constrained to the reactant and product regions.
Furthermore, recent developments in path integral Monte Carlo methods and the availability of massively parallel computers suggests that a double-ended formulation of TPS can be made tractable.

In the present work, we consider the issue of sampling double-ended trajectories for diffusive dynamics.  Going beyond the coarse-graining efforts of Elber and coworkers,\cite{Cze90, Ole96}
we seek the development of an efficient, ergodic sampler of the transition path ensemble.
We show in Section II that the transition path ensemble for diffusive dynamics can be expressed using a \textit{symmetric Onsager-Machlup formula}.  In Section III, we show that discretization of this formula gives rise to a symmetric numerical integrator that requires only first derivatives of the potential energy function and exhibits second order accuracy.  We draw contrast with the standard Onsager-Machlup formula,\cite{Ons53} the discretization of which gives rise to the stochastic Euler integrator that is only first order in accuracy.
In Section IV, we address the path integral Monte Carlo sampling problem that accompanies a double-ended TPS approach.  We begin by establishing the need for fast sampling\cite{Pre05} the fine features of the stochastic process.  We also argue that additional methods are needed to efficiently sample the coarse features of trajectories that are long in time.  We then introduce the \textit{sliding and sampling} algorithm  that alleviates this coarse-feature sampling problem and gives rise to a natural parallelization scheme.  In Section V,  we utilize the S\&S algorithm to sample the ensemble of reactive trajectories for the low-temperature interconversion of the 38-atom Lennard-Jones cluster between its octahedral and icosahedral minima.

\section{The symmetric Onsager-Machlup formula}

The time-evolution of relatively large particles in a highly viscous thermal bath is described by overdamped stochastic dynamics.   In one dimension, these dynamics, which are also known as Brownian or diffusive dynamics, are governed by the law
\begin{equation}
\label{eq:diffusiveEOM}
\dot{x_t} +  \gamma^{-1}V'(x_t) = \dot{W_t}.
\end{equation}
Here, $x_t$ and $\dot{x_t}$ the particle position and velocity at time $t$, and $V(x_t)$ is the external potential energy function.  
The quantity $\gamma$ is the friction coefficient, which is related to the diffusion coefficient ($D=1/(\beta\gamma)$) via the reciprocal temperature $\beta$.  
The white noise $\dot{W_t}$ is formally expressed as the time derivative of a Wiener process, or Brownian motion, of variance $2Dt$.
Extension of Eq.~(\ref{eq:diffusiveEOM}), and the rest of this presentation, to higher dimensions is trivial. 

The object of primary interest in this study is the distribution of transition paths,\cite{TPSrev1,TPSrev2}
\begin{equation}
\label{eq:tpsdist}
P_{AB}(\textbf{x};t)=I_A(x_0)P(\textbf{x};t)I_B(x_n),
\end{equation}
where $P(\textbf{x};t)$ is the equilibrium distribution of  trajectories that obey Eq. (\ref{eq:diffusiveEOM}) and pass through points $\textbf{x}=\{x_0,x_1,\ldots,x_n\}$ in subsequent timesteps of $\Delta t=t/n$.  $I_A(x)$ is the indicator function for the subset of configuration space associate with the reactant species, and $I_B(x)$ is the corresponding product indicator function.
From the Markov property of diffusive dynamics, it follows that in the canonical ensemble
\begin{equation}
\label{eq:markov}
P(\textbf{x};t) = e^{-\beta V(x_0)}\prod_{k=1}^n K(x_k, x_{k-1};\Delta t),
\end{equation}
where $K(x', x;t)$ is the conditional probability that the particle arrives at point $x'$, given that it started at point $x$ some time $t$ earlier.  

$K(x', x;t)$ obeys the Smoluchowski equation, 
\begin{equation}
\label{eq:smolu}
\frac{\partial}{\partial t} K(x', x;t) = D \frac{\partial}{\partial x'} \left\{\left[\frac{\partial}{\partial x'} + \beta V'(x')\right]K(x',x;t)\right\},
\end{equation}
which is the partial differential equation that corresponds to Eq.~(\ref{eq:diffusiveEOM}).
However, by reweighting the conditional probability, 
\begin{equation}
\label{eq:reweight}
G(x',x;t) = e^{+\beta [V(x')-V(x)]/2} K(x',x;t),
\end{equation}
Eq.~(\ref{eq:smolu}) is cast in the self adjoint form of the Bloch equation, 
\begin{equation}
\label{eq:bloch}
\frac{\partial}{\partial t} G(x',x;t) = \left[D \frac{\partial^2}{\partial x'^2} - V_{\mathrm{eff}}(x')\right]G(x',x;t),
\end{equation}
with
\begin{equation}
\label{eq:Veff}
V_{\mathrm{eff}}(x) = D\left[\frac{\beta^2 }{4} V'(x)^2 - \frac{\beta}{2}V''(x)\right].
\end{equation} 
The solution for the conditional probability is then immediately known from the Feynman-Kac solution for the Green's function of the Bloch equation,\cite{Kac51} 
\begin{eqnarray}
\label{eq:FKeq} \nonumber
G(x',x;t) = \mathbb{E}_{x,x'}^{(2Dt)} \exp\left\{-\frac{\beta^2 D}{4}\int_0^t V'(W_u)^2 du \right. \\ + \left. \frac{D\beta}{2}\int_{0}^t V''(W_u)  du\right\},
\end{eqnarray}
with the symbol $\mathbb{E}_{x,x'}^{(2Dt)}$ meaning an average over all Brownian paths of variance $2Dt$ that start at $x$ and end up at $x'$ in time $t$.  

In practice, it will be useful to eliminate the evaluation of second derivatives of the potential function (Eq.~(\ref{eq:FKeq})).  We thus perform a final manipulation. 
Using that
\[
\lim_{\Delta t_k\rightarrow 0}(W_{t_{k+1}} - W_{t_k})^2\rightarrow
 \mathbb{E}^{(2Dt)}(W_{t_{k+1}} - W_{t_k})^2=2D\Delta t_k,
\]
we consider the difference between forward and backward Ito integrals, obtaining
\begin{widetext}
\begin{eqnarray}
\label{eq:itomanip} \nonumber 
\int_{0}^t V'( W_u) \cdot d^bW_u - \int_{0}^t V'( W_u) \cdot d^fW_u  = \lim_{\Delta t \to 0} \sum_{k = 0}^{n-1}  \left[ V'( W_{t_{k+1}})-V'(W_{t_k})\right](W_{t_{k+1}} - W_{t_k}) \\ = \lim_{\Delta t \to 0} \sum_{k = 0}^{n-1}   V'' (W_{t_k})(W_{t_{k+1}} - W_{t_k})^2 = 2D\lim_{\Delta t \to 0} \sum_{k = 0}^{n-1}   V'' (W_{t_k})\Delta{t_k} = 2D\int_0^t V''(W_u)du.
\end{eqnarray}
\end{widetext}
Substituting the last equality in the Eq.~(\ref{eq:FKeq}), we finally obtain the \emph{symmetric Onsager-Machlup} formula
\begin{eqnarray}
\label{eq:SOM} \nonumber
G(x',x;t) = \mathbb{E}_{x,x'}^{(2Dt)} \exp\left\{-\frac{\beta^2 D}{4}\int_0^t V'(W_u)^2 du \right. \\ + \left. \frac{\beta}{4}\int_{0}^t V'( W_u) \cdot (d^b - d^f)W_u \right\}.
\end{eqnarray}

The symmetric Onsager-Machlup formula in Eq.~(\ref{eq:SOM}), which to our knowledge is an original contribution of this work, is so called because of its clear relationship to the standard Onsager-Machlup formula,\cite{Ons53} 
\begin{eqnarray}
\label{eq:OM} \nonumber
K(x',x;t) = \mathbb{E}_{x,x'}^{(2Dt)} \exp\left\{-\frac{\beta^2D}{4}\int_0^t V'(W_u)^2 du \right. \\ - \left. \frac{\beta}{2}\int_{0}^t V'(W_u) \cdot d^fW_u\right\}.
\end{eqnarray}
However, unlike Eq.~(\ref{eq:OM}), Eq.~(\ref{eq:SOM}) is symmetric in $x$ and $x'$.  (That Eq.~(\ref{eq:SOM}) is symmetric is obvious from the inspection of Eq.~(\ref{eq:itomanip}); that Eq.~(\ref{eq:OM}) is not symmetric follows from its dependence on a forward Ito integral and the lack of equality of forward and backward Ito integrals.)
Although the two solutions are formally equivalent, it will be shown in the next section that the symmetric formula gives rise to more stable numerical implementation.  

Before proceeding, we note that the Feynman-Kac equation is also symmetric in $x$ and $x'$ and promises good numerical stability, but we will currently avoid evaluating second derivatives of the potential energy function and reserve exploration of this avenue for a later date.

\section{Second order discretization of the symmetric Onsager-Machlup formula}
Discretization of the standard Onsager-Machlup formula gives rise to Euler's integrator,\cite{Ons53}  a first-order integration scheme.  In this section, we will show that discretization of the symmetric Onsager-Machlup formula gives rise to a more stable, second-order integration scheme. 

A symmetric discretization can be obtained from the symmetric Onsager-Machlup formula  by utilizing the finite difference appearing in Eq.~(\ref{eq:itomanip}) for the difference of Ito integrals and the trapezoidal quadrature technique for the Riemann integral. Thus, letting $t_k = kt/n$ for all $0 \leq k \leq n$, $w_k = t/n$ for all $1 \leq k \leq n-1$, and $w_0 = w_n = t/(2n)$, we obtain the approximation
\begin{eqnarray}
\label{eq:I24} \nonumber
G(x,x';t) \approx\mathbb{E}_{x,x'}^{(2Dt)}  \exp\left\{-\frac{\beta^2 D}{4}\sum_{k=0}^n w_k V'(W_{t_k})^2  \right. \\ + \left. \frac{\beta}{4} \sum_{k = 1}^n [V'( W_{t_k}) - V'( W_{t_{k-1}})](W_{t_k} - W_{t_{k-1}}) \right\}. 
\end{eqnarray}
Since the joint distribution of the Wiener process is known explicitly, the right-hand side can be reduced to a finite-dimensional integral as follows. Define the quantity
\begin{equation}
\label{eq:I25}
p_{\alpha^2}(x,x') = \frac{1}{\sqrt{2\pi \alpha^2}} e^{-(x'-x)^2/(2\alpha^2)},
\end{equation}
and let
\begin{equation}
\label{eq:I26}
\sigma^2 = 2Dt.
\end{equation}
Then, setting $x_0 = x$ and $x_n = x'$, 
\begin{eqnarray}
\label{eq:I27} \nonumber
G(x,x';t) \approx \int_{\mathbb{R}}dx_1\cdots\int_{\mathbb{R}}dx_{n-1} p_{\sigma^2/n}(x,x_1)p_{\sigma^2/n}(x_1,x_2)  \\  \cdots p_{\sigma^2/n}(x_{n-1},x') \exp\left\{-\frac{\beta^2 D}{4}\sum_{k=0}^n w_k V'(x_k)^2  \right. \quad \\ + \left. \frac{\beta}{4} \sum_{k = 1}^n [V'( x_k) - V'(x_{k-1})](x_k - x_{k-1}) \right\}. \nonumber
\end{eqnarray}

Finally, the right-hand side of Eq.~(\ref{eq:I27}) can be expressed in a way that is more familiar to those working with imaginary-time path integrals. $G(x,x';t)$ has the form of a Lie-Trotter product 
\begin{eqnarray}
\label{eq:I28} \nonumber
G(x,x';t) \approx \int_{\mathbb{R}}dx_1\cdots\int_{\mathbb{R}}dx_{n-1} G_0(x,x_1;t/n)\\ \times G_0(x_1,x_2;t/n) \ldots G_0(x_{n-1},x';t/n)
\end{eqnarray}
obtained from the ``short-time'' approximation (remember that $\sigma^2 = 2Dt$)
\begin{eqnarray}
\label{eq:I29} \nonumber
G_0(x,x';t) = p_{\sigma^2}(x,x') \exp\left\{-\frac{(\beta \sigma)^2}{8}\frac{ V'(x)^2 + V'(x')^2}{2}  \right. \quad \\ + \left. \frac{\beta}{4} [V'(x') - V'(x)](x' - x) \right\}. \qquad
\end{eqnarray}
Eqs.~(\ref{eq:tpsdist}), (\ref{eq:markov}), (\ref{eq:reweight}), and (\ref{eq:I29}) fully specify the discretized distribution of transition paths employed in this study,
\begin{equation}
\label{eq:I21}
I_A(x_0)e^{-\frac{\beta}{2} V(x_0)} \left[\prod_{k = 1}^n  G_0(x_{k}, x_{k-1}; \Delta t_{k})\right] I_B(x_n)e^{-\frac{\beta}{2} V(x_n)}.
\end{equation}


In considering the symmetric formula, we have been motivated not only by the need to ensure detailed balance, but also by the observation of Suzuki that the order of convergence of a short-time approximation of the form $e^{-tA}e^{-tB}$ (which is generally one) is improved when the symmetric form $e^{-tA/2}e^{-tB}e^{-tA/2}$ is utilized (to the value of two, which is also the maximal value for such products; a summary of Suzuki's findings can be found in Ref.~\onlinecite{Pre04}). A more general theory, which is also valid outside the domain of applicability of the Trotter theorem, has been designed by one of us.\cite{Pre04} That theory can be utilized both to verify the convergence order of a short-time approximation as well as to design improved ones.\cite{Pre06a} It also allows one to design optimal propagation techniques useful for grid implementations.\cite{Pre06b} 

We shall utilize such a grid technique to verify numerically that the convergence order of the short-time approximation given by Eq.~(\ref{eq:I29}) is two. The grid technique is a sparse version of  numerical matrix multiplication. The choice of grid and propagation technique depends on the order of convergence that is desired. For example, if the convergence order of a spatially-continuous short-time approximation is two, then a grid technique of order two preserves the order of convergence, whereas a grid technique of order greater or equal to three preserves not only the order but also the convergence constant. We shall therefore utilize the fourth-order grid from Ref.~\onlinecite{Pre06b} to verify the convergence order. Once the order is verified and found not to exceed two, one can utilize the second order grid technique for actual computations on a grid. The lower the order of the grid is, the better the sparsity and the speed of the propagation scheme. Grid techniques are generally limited to low dimensional systems. 

The physical system considered is taken from Ref.~\onlinecite{Del98a}. It is a two-dimensional two-channel problem described by the potential
\begin{eqnarray}
\label{eq:I30} \nonumber
V(x,y) = (4(1-x^2-y^2)^2 + 2(x^2-2)^2  + ((x+y)^2-1)^2 \\ + ((x-y)^2-1)^2-2)/6. \qquad
\end{eqnarray}
In reduced units, the friction coefficient is $\gamma = 3$, the inverse temperature is $\beta = 8$, whereas the total propagation time is set to $t=60$, twice the value utilized in the original reference. 

We shall verify the ability of the short-time approximation to preserve the Boltzmann distribution. Using Eq.~(\ref{eq:I21}), if we let $I_A(x) = I_B(x) = 1$ and integrate over all coordinates, we should ideally obtain the value
\[
Z(\beta) = \int_{\mathbb{R}}e^{-\beta V(x)}dx.
\]  
This is the case for the exact Green's function $G(x,x';t)$, which leaves the distribution $e^{-\beta V(x)/2}$ invariant. 
If the short-time approximation $G_0(x,x';t)$ is utilized instead, then the result after propagation, denoted here by $Z_n(\beta)$, is normally different from $Z(\beta)$. It converges to $Z(\beta)$ in the limit $n \to \infty$.

\begin{figure}[!tbp] 
      \includegraphics[angle=270,width=8.5cm,clip=t]{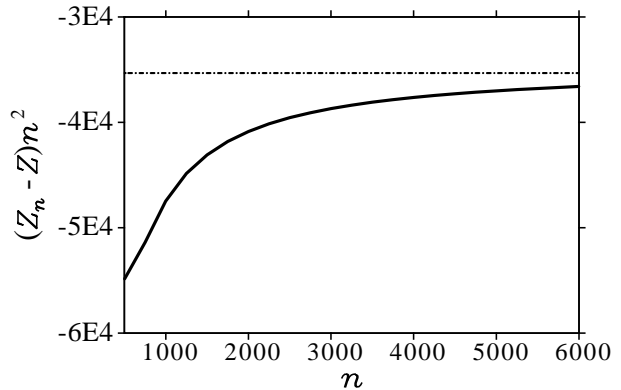} 
 \caption[sqr]
{\label{Fig:1}
The instantaneous convergence constant for the partition function as a function of the number of time slices. The guiding line is the convergence constant determined for $n = 16,000$.   
}
\end{figure}

We want to establish how fast this convergence is. An easy way is to verify that the difference $n^2[Z_n(\beta) - Z(\beta)]$ converges to a constant different from zero (the convergence constant). As this is apparent from Fig.~\ref{Fig:1}, second-order convergence is numerically established. Because of the long times utilized, the number of path variables necessary for an accurate treatment of rare events is large. For our example, a number of slices equal to $n = 2048$ produces a relative accuracy  of $-2.3\%$ for the partition function. 

\section{Path Integral Monte Carlo Sampling}

The distribution of trajectories defined by Eq.~(\ref{eq:I21}) poses  a path integral Monte Carlo (PIMC) sampling problem.  We can thus employ existing PIMC sampling techniques as a starting point for sampling the ensemble of diffusive transition paths.\cite{Cep95,Pre05}  However, the goal of sampling trajectories that are \textit{long in time} gives rise to challenges that have not been fully addressed by current methods.  These issues, and the techniques that we propose for overcoming them, are discussed in the current section.

\subsection{$\quad$Sampling Fine Features of the Path: $\quad\quad$ The Fast Sampling Algorithm}

Define $l_V$ to be the characteristic lengthscale of the potential energy function, namely the lengthscale below which the forces arising from the potential function are slowly varying.
Furthermore, define $\tau_V~\equiv~l_V^2/2D$, the typical time required for a free particle to diffuse a distance $l_V$.
If we then consider a segment of length $\tau$ from any diffusive trajectory such that $\tau\ll\tau_V$,
we see immediately from Eq. (\ref{eq:I29}) that the distribution for this path segment is governed almost entirely by the Gaussian (free-particle) term.  It follows that diffusive trajectories experience a nearly drift-free Brownian motion on short timescales, which implies that such trajectories are composed of nearly independent infinitesimal increments.

As is well known,\cite{Cep95,Pre05}
this near-independence of the infinitesimal increments must be respected in a sampling algorithm if the usual path integral stiffness problem is to be avoided.  For example, sampling the path distribution with individual time-slice moves gives rise to an artificial correlation between the increments (often blamed on the ``harmonic springs'' in the free particle term) which causes poor computational scaling with the number of time-slices $n$.\cite{Cep95}

A solution to this problem is to use collective Monte Carlo moves for the time-slices, or multi-slice moves.\cite{Cep95}
A general strategy for doing so is to express the diffusive trajectories using a random series representation, namely an infinite series of specialized functions,\cite{Pre02}
\begin{equation}
x+(x'-x)u+\sqrt{2Dt}\sum_{i=1}^{\infty}a_i\Lambda_i(u),
\label{randomseries}
\end{equation}
where $\quad0\leq u\leq1$.
The conditions for these specialized functions are dictated by the Ito-Nisio theorem.\cite{ItoNis}
Permissible functions include the well known
Wiener-Fourier (sine function) series,\cite{Dol84,Coa86,Pre02}
 as well as the L\'evy-Ciesielski series\cite{Pre05} that will be discussed shortly.

Random series representations have several important properties.
In particular, the $\Lambda_i(u)$ are constructed such that their amplitudes diminish with increasing index $i$; the tail of the random series thus becomes nearly decoupled from the potential function and describes the fine features of the path. 
Furthermore, in the special case of $V'(x)=const$, 
Eq. (\ref{randomseries})  yields the correct distribution
of trajectories upon letting the $a_i$ be a set of independent identically distributed standard normal variables.
Together, these properties ensure that the path variables in the random series representation become nearly decoupled in describing the fine features of the path.  That is, direct Monte Carlo sampling of the $a_i$ generates multi-slice moves that avoid introducing artificial correlation between the infinitesimal increments of the path.

The L\'evy-Ciesielski random series (LC)  representation is
especially amenable to efficient path sampling.  In particular, the recently developed Fast Sampling Algorithm (FSA) utilizes the LC representation to provide, for a fixed trajectory time, arbitrarily high accuracy in sampling the fine features of the path at a computational scaling of only  $n\ \textrm{log}_2 n$ in the number of path variables.\cite{Pre05}  
We briefly review the L\'evy-Ciesielski random series representation and the fast sampling algorithm, which are employed in the numerical examples throughout this study, and we refer the reader to Ref.~\onlinecite{Pre05} for a more complete discussion.

In the LC representation, the path is expanded in the basis of Schauder functions,
\begin{equation}
F_{k,j}(u)=2^{-(k-1)/2}F_{1,1}(2^{k-1}u-j+1),
\label{Fkj}
\end{equation}
where $k\ge1$, $1\le j\le 2^{k-1}$, and
\begin{equation}
F_{1,1}(u)=\left\{\begin{array}{ll}
	u& u\in (0,1/2]\\
	1-u & u\in (1/2,1)\\
	0 & \textrm{otherwise.}\end{array}\right. 
\end{equation}
These functions, which are illustrated in Fig. 2,
are organized into $k$-indexed layers that describe increasingly coarse features of the path.  Each layer contains a set of $2^{k-1}$ functions which are only non-zero on the disjoint, open intervals $(u_{k,j-1},u_{k,j})$, where $1\le j\le 2^{k-1}$ and $u_{k,j}=j2^{-(k-1)}$.  
Because of this property, we know that only one $j$-indexed function within a layer contributes to the expansion, such that
\begin{equation}
\sum_{j=1}^{2^{k-1}}b_jF_{k,j}(u) = b_{[2^{k-1}u]+1}F_{k,[2^{k-1}u]+1}(u)
\end{equation}
for any sequence of numbers $b_1,b_2,\ldots,b_{2^{k-1}}$.  (Here, [x] denotes the largest integer smaller than or equal to x, and for $u=1$, the quantities $b_{2^{k-1}+1}$ and $F_{k,2^{k-1}+1}(1)$ are defined to be zero.)
We thus have the following LC representation of the path,
\begin{equation}
x+(x'-x)u+\sqrt{2Dt}\sum_{k=1}^{\infty} a_{k,[2^{k-1}u]+1}F_{k,[2^{k-1}u]+1}(u).
\label{LCseries}
\end{equation}

\begin{figure}[!tbp] 
   \includegraphics[angle=270,width=8.5cm,clip=t]{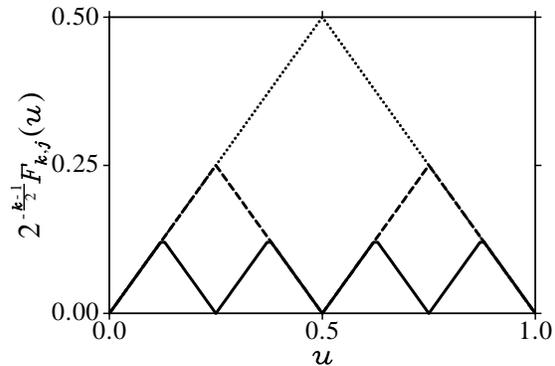} 
 \caption[sqr]
{\label{Fig:2}
Plot of renormalized Schauder functions for the first three layers.
}
\end{figure}

Using the LC representation, we sample the path variables $a_{k,j}$ according to the distribution defined in Eq.~(\ref{eq:I21}).   In doing so, we employ the Fast Sampling Algorithm (FSA), which prescribes that each path variable be sequentially updated using the Metropolis Monte Carlo algorithm.\cite{Pre05}  
The FSA exploits two key properties of the LC representation to enables the efficient sampling of large numbers of time-slices.

Suppose that the path is described using $n+1$ time-slices (including endpoints), where $n =2^{k'}$,
 and consider a particular LC path variable $a_{k,j}$ such that $k>1$.  
As is clear from the preceding paragraph, changing this variable involves moving only $n_k-1$ time-slices, where $n_k=2^{k'-k+1}$.  A beautiful consequence of this fact is that, for path integrals of the Lie-Trotter product form, the change in the path integral action accompanying a change in $a_{k,j}$ can be obtained at only a fraction ($\approx 2^{1-k}$) of the  computational cost of evaluating the action for the total path. (This is an unusual feature for a random series representation.  For example, changing a single path variable in the sine function series requires reevaluating the action for the entire path.)  We thus arrive at the key properties of the LC representation:
(1) In sampling a Lie-Trotter product form,
the path variables that belong to the same layer are strictly independent.
(2) The cost of computing the changes in path integral action for a sequential sweep through all $n$ path variables in the LC representation is only logarithmically greater (a factor $k'$ times) the cost of doing so in the time-slice representation. 
Since it is proven that optimal statistical statistical efficiency of independent variables is obtained by a sequential Metropolis Monte Carlo scheme,\cite{Pre05}
 it is clear that the FSA exploits the first of these properties.
 The FSA also exploits the second property, which ensures that path variables can be introduced into the tail of the random series at a computational cost that scales as only $n\ \textrm{log}_2 n$,
 to ideally solve the problem of sampling the fine features of the path.
 
What is so special about the symmetric  Onsager-Machlup formula that requires utilization of the FSA? The main source of concern is the appearance of Ito integrals in the expression for the action. The difference of backward and forward Ito integrals appearing in Eq.~(\ref{eq:SOM}) would vanish if the path were a smooth function. In fact, if $\phi(u)$ is piecewise smooth, then both the forward and the backward Riemann sums define a same Riemann-Stieltjes integral, that is, 
\begin{eqnarray}
\label{eq:fs4} \nonumber
\int_{0}^1 V'(W_{ut})d^f\phi(u) = \int_{0}^1 V'(W_{ut})d^b\phi(u) \\ = \int_0^1 V'(W_{ut})\dot{\phi}(u)du.
\end{eqnarray}
Hence, the Ito integral in Eq.~(\ref{eq:SOM}) would vanish instead of converging to a quantity involving the Laplacian of the potential, as shown by Eq.~(\ref{eq:itomanip}). It is only the very fine details (the so-called noise) that prevents the difference of Ito integrals from vanishing. 

If, for example, a Lennard-Jones cluster at low temperature starts to suddenly exhibit unexplained liquid-like behavior during a path-integral simulation, the researcher should recall that there are two main reasons that the difference of Ito integrals  vanishes accidentally: i) \emph{inadequate sampling of the fine details} and ii) \emph{a number of time slices that is too small for the chosen transit time} (not enough fine details). If the difference of backward and forward Ito integrals vanishes or is too small, then the distribution of paths is controlled only by the square-norm of the gradient. The distribution tends to accumulate around the stationary points of the potential, where the gradient vanishes. However, the simulation has no way to discern whether these stationary points are local minima, saddle points, or local maxima. It is worth mentioning that the explicit inclusion of the Laplacian term through a numerically better-behaved Riemann integral makes the Feynman-Kac formula less dependent upon the quality of sampling. 

\subsection{Two Kinds of Sampling Stiffness}

A central observation of the current paper is the following: Although random series representations \textit{avoid} stiffness in sampling the fine features of the path, they \textit{introduce} stiffness in sampling its coarse features. 
By using multi-slice moves, random series representations introduce correlations between distant segments of the path.
That is, changing a single path variable in a random series representation generally involves moving a large number of time-slices along the entire length of the path (imagine, for example, changing a coefficient in the sine function expansion).
This is not a problem in sampling the fine features of the path, because the amplitude of the motion is small in comparison to $l_V$ - too small to ``feel'' the potential function. 
But the introduced correlation is a severe problem in sampling the coarse features of long paths.  If the trajectories are long in time ($t\gg \tau_V$), then well-separated segments of the path will in fact be almost entirely uncorrelated - and should be sampled accordingly.  
By using a representation that artificially correlates the well-separated segments - and attempts to move them over distances that are comparable to $l_V$ -  we introduce stiffness.

Stiffness in sampling the coarse features of the path is an obvious problem in the context of reactive trajectories, which travel from a reactant region to a distinct product region - and thus involve motion on lengthscales that are long compared to $l_V$.  However, we note that it is an inevitable problem for \textit{any} path integral Monte Carlo application that utilizes a random series representation and involves long times (including imaginary times).  
Regardless of the context,
we would expect to encounter stiffness whenever dealing with potentials that have a characteristic lengthscale that is short in comparison to the lengthscale for free particle motion ($(2Dt)^{1/2}$ in the current context, $(\beta\hbar^2/m)^{1/2}$ for imaginary time path integrals).

In the following example, we illustrate how stiffness associated with sampling the coarse features of the path appears in practice.
 We return to the two-dimensional, two-channel problem considered in the previous section. 
Using reduced units, we set the friction coefficient $\gamma = 3$, the inverse temperature $\beta = 8$, and the trajectory time $t=60$.  We describe the path using $n=2048$ time-slices, which corresponds to employing $k'=11$ layers in the LC representation.  The endpoints of the trajectory are held fixed at $(x=\pm 1,y=0)$.  (Of course, additionally sampling the endpoints is trivial, but it is also superfluous when a sufficiently long $t$ is employed.)  Employing the FSA, the LC path variables are updated sequentially according to the Metropolis algorithm.  
We propose new path variables according to 
\begin{equation}
a_{k,j} = a^0_{k,j} + \Delta a_{k,j} \xi,
\end{equation}
where $\xi$ has probability density $2^{-1}(1+x^2)^{-3/2}$. Random 
numbers having this distribution can be generated via the identity $\xi = 
(\zeta-0.5) / (\zeta(1-\zeta))^{-1/2}$, with $\zeta$  uniform  on $[0,1]$.  
 The widths of the proposal distributions $\Delta a_{k,j}$ are restricted to be the same for all path variables within a given $k$-indexed layer, and the numerical values for these widths are tuned in an equilibration phase to yield approximately $40\%$ acceptance.  

\begin{table}
 \caption[sqr]
{\label{tab:disp} Proposal distribution widths for the Metropolis Monte Carlo moves in the two-dimensional example.$^a$}
\begin{tabular}{c c c c}
\hline
 $\quad k\quad$& $\quad n_k\quad$ &$\quad\Delta a_{k,j}\quad$ &  $\quad\Delta x\quad$\\
\hline
1 & 2048 &  0.031 & 0.070 \\
2 &  1024&   0.060 & 0.095\\
3 &  512&   0.122 & 0.136\\
4 &  256&   0.248 & 0.196\\
5 &  128&   0.485 & 0.271 \\
6 &  64&   0.831 & 0.328\\
7 &  32&    1.225 & 0.342\\
8 &  16&    1.438 & 0.284\\
9 &  8&    1.545 & 0.216\\
10 &  4&    1.576 & 0.156\\
11 &  2&    1.589 & 0.111\\
\hline
\end{tabular}\\
$^a$ $k$ is the layer index for the LC representation of the path, $n_k-1$ is the number of time-slices moved by changing the path variable $a_{j,k}$,  $\Delta a_{k,j}$ is the equilibrated path variable distribution width, and $\Delta x$ 
 is the corresponding distance in coordinate space.
\end{table}

The results for the equilibrated proposal distribution widths $\Delta a_{k,j}$ are presented in Table \ref{tab:disp}.
We also provide the corresponding distances in coordinates space 
$\Delta x~=~(Dt)^{1/2}2^{-k/2}\Delta a_{k,j}$, 
obtained using Eqs. (\ref{Fkj}) and (\ref{LCseries}).

For layers $k=9-11$, the $\Delta a_{k,j}$ reach a plateau, indicating
 that the tail of the random series expansion has decoupled from the potential energy surface and that the simulation is converged with respect to the number of path variables.
In this tail of the expansion, we see from the $\Delta x$ values that subsequent layers sample increasingly fine features of the path.

However, for $k<9$, the $\Delta a_{k,j}$ decrease with decreasing $k$, eventually falling off exponentially for the most coarse layers.  This behavior indicates that path variables are no longer sampled from their free particle distributions and increasingly feel the presence of the potential energy surface.  The onset of this decrease in proposal widths at $k=8$ is reasonable, given that the characteristic distance traveled by a free particle in $n_k=16$ time-slices is about $0.2$ - a distance that is commensurate with the finest details of the potential surface.

For layers $k=5-8$ in Table \ref{tab:disp}, we note that although the distribution widths $\Delta a_{k,j}$ decrease, the corresponding distances moved by the time-slices remains large and relatively constant.
This feature arises because LC layers with lower $k$ indices correspond to larger lengthscales in configuration space.  Despite the decrease in the proposal distribution widths, we see that the LC representation is still efficiently sampling the time-slices associated with these middle layers.

However, for $k<5$ we see an exponential decrease in both the proposal distribution widths and the corresponding time-slice moves in configuration space.  This follows from the attempt of ever-coarser moves in the a random series representation and inevitably leads to coarse-feature sampling stiffness.

\subsection{$\quad$Sampling Coarse Features of the Path: $\quad\quad$ The Sliding and Sampling Algorithm}

In the preceding subsection, we recognized that random series representations, by utilizing multi-slice moves, introduce stiffness in the Monte Carlo sampling of the coarse features of long paths.  We now offer a simple solution to this problem.
We propose an algorithm in which the total path is divided into shorter fragments. 
Keeping its endpoints fixed, each fragment is represented with a random series and independently sampled using Metropolis Monte Carlo.  The Metropolis-evolved fragment configurations are then used to reconstruct the total path - such that all time-slices other than the fragment endpoints have generally moved. Finally, to ensure ergodicity, the total path is re-divided into a different set of fragments and the process is repeated.

How does this approach avoid both coarse-feature and fine-feature sampling stiffness?  Clearly, by using a random series representation for the fragments, we avoid the fine-feature sampling stiffness.  
Moreover, by fragmenting the path, we achieve ergodic sampling without employing the multi-slice Monte Carlo moves that suffer most extremely from coarse-feature stiffness.

This algorithm essentially generates a rejectionless diffusive dynamics for the fragment endpoints, which in turn define the domains for independent, Metropolis sampling.  Since diffusive dynamics is itself just a sampling device, we are, in a sense, sampling the sampler.

Our specific implementation of this idea utilizes the LC representation.  Presume that the total path of time $t$ is uniformly discretized into $n+1$ time-slices,
\begin{equation}
\textbf{x}=(x_0,x_1,\ldots,x_{n-1},x_n).
\end{equation}
We fragment the path into $f$ segments containing $n_f+1$ time-slices (including endpoints), where $n_f=2^{k_f}$
and $fn_f<n$.
Each fragment can then be represented using an LC series with $k_f$ layers.
Upon requiring that the fragments form consecutive segments of the total path (i.e. share endpoints), there are $n+1-fn_f$ possible ways to fragment the path.
For example, if $n=10$, $f=2$, and $n_f=4$, the three possibilities are
\begin{equation}
\textbf{x}=
\left\{
\begin{array}{l}
	(x_0,x_1,x_2,x_3,x_4),(x_4,x_5,x_6,x_7,x_8),[x_8,x_9,x_{10}]\\
	\left[x_0,x_1\right],(x_1,x_2,x_3,x_4,x_5),(x_5,x_6,x_7,x_8,x_9),[x_9,x_{10}]\\
	 \left[x_0,x_1,x_2\right],(x_2,x_3,x_4,x_5,x_6),(x_6,x_7,x_8,x_9,x_{10}),\end{array}
	\right.\nonumber
\end{equation}
where the segments included in parentheses are the fragments, and those included in square brackets are the leftover ends of the trajectory.
When the path is re-divided between fragment sampling steps, we randomly choose from among these various possibilities, effectively sliding the consecutive fragment sampling domains along the stationary path.
Each fragment sampling step involves a single sweep through the fragment LC path variables using the FSA.  We call this procedure \textit{sliding} and \textit{sampling} (S\&S). 

\begin{figure}[!tbp] 
\includegraphics[angle=270,width=7.5cm]{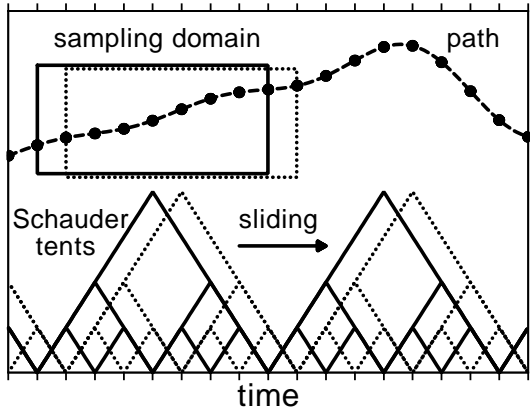} 
\caption[sqr]
{\label{sliding}
Schematic illustration of a sliding move.  Using Schauder tents, the LC representation for the path is shown in solid lines before sliding and in dotted lines after sliding.  The segment of the path that corresponds to a particular fragment, before and after sliding, is shown using the solid and dotted boxes, respectively.
}
\end{figure}

The sliding move is illustrated schematically in Fig. \ref{sliding}.  The top of the figure shows a segment of the total path, which extends out of view in both directions, as a function of time.  
At the bottom, in solid lines, we see the LC representation for a particular fragmentation of this path, using lengths of $n_f=8$.  The LC layers are illustrated in the usual way with a nested structure of ``Schauder tents.''  The solid box indicates the segment of the path that corresponds to a particular fragment.
In dotted lines, the figure shows how the path is re-fragmented in a sliding move.
Although the path is unmoved, its representation in the LC path variables changes as we slide the fragment endpoints to the right by one time-slice.  The dotted box illustrates how the domain over which the path is sampled by the corresponding fragment slides as well.

In explaining the S\&S algorithm above, we did not prescribe Metropolis sampling for the ends of the total path. 
  This does not pose a formal problem for ergodicity, but to provide optimal sampling of the path, it is best not to unduly neglect any time-slices.  In practice, we also represent these ends in terms of LC path variables which are subjected to Metropolis sampling. 

An important feature of the S\&S algorithm is that it provides symmetric sampling for all of the time-slices in the path, despite the intrinsic asymmetry of the LC representation.
To illustrate this, consider sampling the entire path in the LC representation without fragmentation.  The time-slice corresponding to the exact middle of the path ($t/2$) is then \textit{only} moved by changing the $a_{1,1}$ LC path variable.  In contrast, the time-slices that directly neighbor the middle time-slice will be moved by changing LC path variables in \textit{any} layer.
This asymmetry of the LC representation causes some path variables to suffer from coarse-feature sampling stiffness more drastically than others, and it limits the efficiency of sampling the entire path to the efficiency of sampling its most coarse (and most stiff) layers.
The S\&S algorithm removes this bottleneck by ensuring that a particular time-slice appears at all positions within a fragment with equal probability.
Each time-slice is thus exposed to the full range of Monte Carlo moves.
Fragmentation, in general, ensures multi-resolution sampling of the time-slices, regardless of the deficiencies of the underlying random series representation.

Another important feature of the S\&S algorithm is that it is naturally parallelizable.
The Markov property of the diffusive dynamics 
 guarantees that the distribution of a particular path fragment  only depends upon the position of its endpoints.  By keeping these endpoints fixed during the sampling step, we ensure that the fragment distributions are statistically independent.  It follows that the task of sampling the individual fragments can be performed, in parallel, by up to $f+2$  (= number of fragments + 2 ends) distributed-memory nodes.
We note that since a sampling sweep over the path variables in a given fragment requires $n_f$ evaluations of the force field, the cost of communication between nodes in the S\&S algorithm is negligible.


A consideration in implementing the S\&S algorithm is the choice of the number of time-slices per fragment.
In principle, to minimize the total expenditure of computer hours, choosing $n_f$ requires balancing the cost of coarse-feature stiffness introduced by the random series representation with the cost of (Gibbs-type) correlation introduced by excessive fragmentation of the path and the corresponding elimination of productive multi-slice moves.  
However, we emphasize that use of the S\&S algorithm with \textit{any} choice of $n_f$ eliminates the bottleneck in sampling efficiency due the asymmetry of the LC representation.
It follows that the maximum computational penalty to be incurred with excessively long fragments is the logarithmic increase in computer time associated with attempting inefficient moves in the stiff coarse layers.  We thus anticipate that the penalty in computational cost for overestimating $n_f$ in the S\&S algorithm is a minor concern.
This issue is explored in the following numerical example.

In Table \ref{tab:2d}, we again consider the two-dimensional, two-channel numerical example.  Using $n=2048$ time-slices, we sample path distribution using the S\&S algorithm with different numbers of fragments, and we consider the performance of the algorithm in terms of computer and wall-clock time.  As a measure of the quality of sampling in each simulation, we count $n_{hops}$, the number of times that the midpoint of the total path crosses barrier in the potential energy surface at $x=0$.  

\begin{table}
 \caption[sqr]
{\label{tab:2d} Sampling efficiency for the two-dimensional example using the parallelized S\&S algorithm with various fragment lengths.$^a$}
\begin{tabular}{c c c c c c}
\hline
$f+1$ & $\ \ n_f\ \ $ & $k_f$& $\frac{n_{hops}}{\textrm{S\&S sweep}}$ &  $\frac{n_{hops}}{\textrm{CPU time}}$  & $\frac{n_{hops}}{\textrm{wall-clock time}}$ \\
\hline
2&   1024 &10 &100   &  100   &  100  \\
4&   512   & 9  & 91    &101      & 168 \\
8&   256   & 8  & 85    &106      & 317 \\
16& 128   & 7  & 67    &96        & 543 \\
32& 64     & 6  & 42    &70        & 765 \\
64& 32     & 5  & 15    &31        & 663 \\
128&16     & 4  &  5    &13        & 578 \\
\hline
\end{tabular}\\
$^a$ $f$ is the number of fragments.  $n_f+1$ is the number of time-slices per fragment. $k_f$ is the number of layers in the LC series used to describe each fragment.  Data in the last three columns are normalized.
\end{table}

The first three columns of the Table \ref{tab:2d} characterize the number and length of the fragments used in the S\&S algorithm.  $f$ is the number of fragments,  $n_f$ is the number of time-slices per fragment, and $k_f$ is the number of layers in the LC representation used to describe each fragment. 
In the fourth column, we show the number of hopping events observed per number of sweeps through the S\&S algorithm (each of which includes a Monte Carlo sweep through the fragment path variables and a sliding move).
The results are normalized with respect to the most coarse fragment.
Clearly, better sampling per sweep is achieved with larger fragments.  
This is primarily due to the fact that increasing the fragment length in the LC representation corresponds to attempting additional (coarse-layer) Monte Carlo moves per S\&S sweep.  An additional source of improvement is that the correlation in sampling the fragment endpoints (i.e. Gibbs-type correlation) decreases with increasing fragment length.

Although improved sampling \textit{per sweep} is obtained with longer fragments, coarse-feature stiffness is realized because these improvements do not fully compensate for the increasing computational cost of performing each sweep.  This is illustrated in the fifth column of Table \ref{tab:2d},
in which the data from column four are scaled by the relative computational costs of performing an S\&S sweep with different fragment lengths.
(To compare with the results for the most coarse fragmentation, each entry in column four is multiplied by the factor $\textrm{log}_2(1024)/\textrm{log}_2(n_f) = 10/k_f$.)
As $n_f$ increases from $16$ to $256$, we observe the improved sampling efficiency that accompanies longer fragments.
However, at approximately $n_f=512$, the computational cost of performing moves in the coarsest layer exceeds the corresponding improvement in the sampling.  Beyond this point, we increasingly suffer from the mild coarse-feature sampling stiffness of the LC representation.

The final column in Table \ref{tab:2d} reveals the parallelizability of the S\&S algorithm.  The number of observed hops per unit of wall-clock time is reported by scaling
the results in column five by the relative number of computational nodes that can by harnessed via parallelization.  (For comparison with the most coarse fragmentation, each entry in column five is multiplied by the factor
$(f+2)/3$.)
These results again reveal a turnover in sampling efficiency - but at much shorter fragmentation.
Better sampling per unit wall-clock time is generally obtained by enlisting the efforts of more computational nodes.  Eventually, this trend breaks down when we lose the benefit of treating the fine-scale features of the path with the random series representation.   Specifically, we see that when we employ fragments with fewer than $64$ time-slices, further parallelization is not beneficial. 

 This last observation is consistent with the $\Delta x$ values reported in Table \ref{tab:disp}.  Both sets of data suggest that, for this particular problem and these particular parameters, the LC representation is very efficient in sampling path variables with fewer than $64$ time-slices.  Fragmenting the path beyond this point diminishes the quality of sampling.
It is convenient that the equilibrated maximum displacements indicate the lower bound for the acceptable number of time-slices per fragment.

Before proceeding, we can draw some practical conclusions.
The coarse-feature stiffness gives rise to an expected turnover in CPU-time efficiency with increasing fragment length.  However, this penalty is very mild (as was also expected), suggesting that little CPU time is wasted by utilizing very long fragments.  Furthermore, it is possible that coarse moves might provide valuable flexibility in the sampling of complicated problems, regardless of their stiffness.
We thus have considerable freedom in choosing an acceptable length for the fragments in the S\&S algorithm.  If one only has access to a small number of CPU nodes, using the S\&S algorithm with long fragments promises ``optimal'' efficiency in terms of computational effort.
On the other hand, if one has access to a large number of CPU nodes, then use of shorter fragments leads to a lower degree of CPU-time efficiency but a vast improvement in wall-clock-time efficiency.
To summarize: Good sampling efficiency can be expected in the S\&S algorithm if the fragments are long enough to avoid fine-feature sampling stiffness.  Beyond this restriction, the final choice of fragment length can be made on the basis of the number of available CPU nodes and the desire to minimize wall-clock time.
Finally, we note that the S\&S algorithm is sufficiently general to incorporate possible improvements in the underlying sampling techniques, as they come available.

\section{Application to the 38-atom Lennard-Jones Cluster}

As a challenging test of our double-ended approach to sampling diffusive trajectories, 
we consider the structural rearrangement of the 38-atom Lennard-Jones cluster (LJ$_{38}$) at low temperature.
Unlike most other small Lennard-Jones clusters, LJ$_{38}$ does not have a global minimum energy structure that exhibits  an icosahedral core.\cite{Wal97,Doy99a,Doy99b,Lea99}
  Instead, its global minimum is a symmetric truncated octahedron with a total binding energy that is only $0.38\%$ lower than the most stable icosahedral structure.\cite{Doy99a}  
  The LJ$_{38}$ potential surface exhibits a double-funnel topology, with one basin of stability exhibiting octahedral-like structures and the other exhibiting icosahedral-like structures.  For temperatures below $T=0.1\ \epsilon/k_B$, the thermal distribution is localized almost entirely in the octahedral basin.  At this temperature, a finite-system solid-solid phase transition (or ``phase change'') occurs in which the distribution shifts to the entropically favored icosahedral basin.\cite{Nei00}  At $T=0.166\ \epsilon/k_B$, a melting transition occurs as the vast number of disordered, liquid-like structures become thermally accessible.\cite{Nei00,Doy98}

Sampling the LJ$_{38}$ cluster is known to be extremely difficult.\cite{Doy99a}
Direct Metropolis Monte Carlo sampling of its low-temperature thermodynamics is not feasible.\cite{Nei00,Cal00}  
 At the low temperatures for which the cluster is solid, thermally accessible regions of configuration space are separated by very large energetic barriers, and the potential surface contains many local minima that pose sampling (and kinetic) traps.  Furthermore, the octahedral basin, and the pathways by which it is accessible, are sufficiently narrow to make tortuous any sampling algorithm's task of equilibrating the two basins.  

 One also expects that the \textit{dynamics} of interconversion between the octahedral and icosahedral basins is extremely frustrated at low temperature.  Our preliminary efforts to sample the ensemble of reactive trajectories using the standard TPS algorithm with shooting and shifting confirms this to be the case.  Even when long trajectories are employed, the system simply can not find the basins of stability; it instead becomes trapped in disordered, high-energy structures.

Given the difficulty of ergodically sampling the configuration space of LJ$_{38}$ at low temperature, it is unlikely that we will be able to fully achieve this feat in path space with the techniques introduced here.
After all, we have only addressed the specific problems associated with sampling long dynamical trajectories - not the intrinsic deficiencies of the Metropolis Monte Carlo algorithm.
However, we can expect to harvest reasonable examples of transition pathways and to achieve locally-ergodic sampling in the space of trajectories.  

We employ the S\&S algorithm to sample the distribution of dynamical trajectories corresponding to interconversion between the octahedral and lowest icosahedral structures of the LJ$_{38}$ cluster.  These trajectories connect permutationally distinct isomers of the octahedral and icosahedral structure.  Specifically, we choose the two isomers that minimize the Euclidian distance between the endpoints of the path at $3.274 \sigma$, as reported by Trygubenko and Wales.\cite{Try04}

The potential energy function, in the Lennard-Jones reduced units employed hereafter, is 
\begin{equation}
V(\textbf{r})=4\sum_{i<j}^{38}\left[r_{ij}^{-12}-r_{ij}^{-6}\right] + V_c(\textbf{r}),
\end{equation}
where $r_{ij}$ is the distance between the $i^{th}$ and $j^{th}$ atoms in the cluster, and $V_c(\textbf{r})$
is the constraining potential used to prevent evaporation of the cluster.  It is defined
\begin{equation}
V_c(\textbf{r})=\sum_{i=1}^{38} v_c(\delta_i),
\end{equation}
where $\delta_i$ is the distance between the $i^{th}$ atom and the center of mass of the cluster, and
\begin{equation}
v_c(\delta)=\left\{\begin{array}{ll}
	0,& \delta \le r_c\\
	\alpha\left(\frac{\delta-r_c}{r_c}\right)^3,& \delta >r_c,\end{array}\right. 
\end{equation}
with $\alpha=5000$ and $r_c = 2.65$.

We sample the ensemble of reactive trajectories at the phase change temperature, $T=0.1$.  We employ a friction coefficient $\gamma=1$ and a total simulation time $t=1$.  
Two simulations are performed using different fragment lengths.  In the first, we employ $f=30$ fragments of length $n_f=128$ time slices and a number of time-slices $n=(f+1)*n_f=3968$.
In the second, we employ $f=14$ fragments of length $n_f=256$ and a number of time-slices $n=3840$.  This small difference in $n$ is not sufficient to alter the accuracy of sampling. 
The simulations were initialized from a straight line trajectory between the minimized icosahedral and octahedral isomers, and these endpoints were held fixed during the simulations.  The simulations were equilibrated for $10^6$ S\&S sweeps, during which the proposal distribution widths were updated to yield approximately $40\%$ acceptance.   They were then continued for an additional $3\times 10^6$ production sweeps.  Trajectories were recorded every $10^4$ sweeps.
  The simulation with $n_f=128$ was run in parallel on $32$ single-processor, distributed-memory nodes, and the simulation with $n_f=256$ employed $16$ nodes.  The total required CPU time, for each simulation, was approximately $10^5$ hours.
  
\begin{table}
 \caption[sqr]
{\label{tab:dispLJ} Proposal distribution widths for the Metropolis Monte Carlo moves in the LJ$_{38}$ application.$^a$}
\begin{tabular}{c c c c}
\hline
 $\quad k\quad$& $\quad n_k+1\quad$ &$\quad\Delta a_{k,j}\quad$ &  $\quad\Delta x\quad$\\
\hline
1  &  256&   0.041   & 0.005\\
2  &  128&   0.086   & 0.007\\
3  &   64&   0.177    & 0.010\\
4  &   32&    0.309   & 0.012\\
5  &   16&    0.546   & 0.015\\
6 &   8&    0.922    & 0.018\\
7 &   4&    0.987    & 0.014\\
8 &   2&    1.161    & 0.011\\
\hline
\end{tabular}\\
$^a$ $k$ is the layer index for the LC representation of the fragment, $n_k$ is the number of time-slices moved by changing the path variable $a_{j,k}$,  $\Delta a_{k,j}$ is the equilibrated path variable distribution width, and $\Delta x$ is the corresponding distance in coordinate space.
\end{table}

In Table \ref{tab:dispLJ}, we report the equilibrated path variable proposal widths for the $n_f=256$ simulation.  The results are nearly identical to those obtained for $n_f=128$, and they have the same features as were discussed for the two-dimensional example in connection with Table \ref{tab:disp}.  In particular, we note a plateau in the $\Delta a_{k,j}$ values for $k>6$ as the path variables become decoupled from the potential energy surface.  The plateau value seen here is smaller than was found in Table \ref{tab:disp} because we are now dealing with three-dimensional, rather than two-dimensional, particles.
For $3\le k\le 6$, the $a_{j,k}$ decrease with decreasing $k$ while the corresponding $\Delta x$ values remain large, indicating that these layers remain productive in sampling the path.  Finally, for $k\le2$, the $\Delta x$ rapidly decrease as the most coarse layers approach the onset of stiffness.
This table suggests that a fragmentation length of at least $n_f=64$ is necessary to avoid the exclusion of LC path variables that substantially contribute to the sampling efficiency.

\begin{figure}[!tbp] 
\includegraphics[angle=270,width=8.5cm]{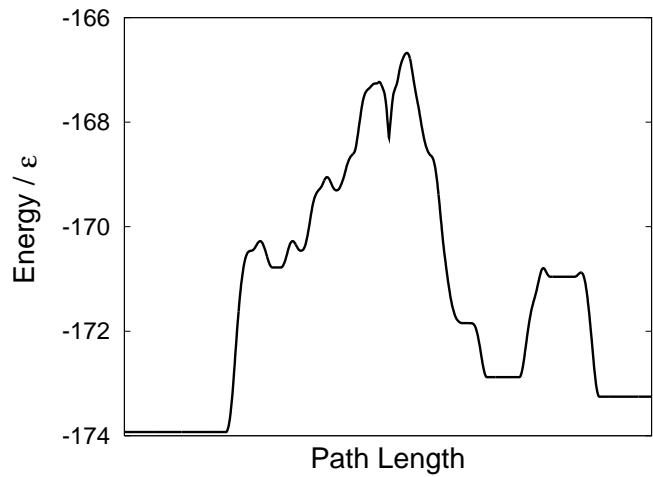} 
\caption[sqr]
{\label{afterquench}
A typical quenched pathway for interconversion between the octahedral and icosahedral LJ$_{38}$ isomers, obtained by sampling with the S\&S algorithm and quenching using the ZTS method.
}
\end{figure}

For the purposes of analysis, the sampled trajectories were quenched on the potential energy surface
 using the zero temperature string (ZTS) method of E, Ren, and Vanden-Eijnden.\cite{E02}
Given an initial path, this method yields a continuous, minimum-energy path (i.e. one that is everywhere tangent to the gradient of the potential energy function and is compose of equidistant images of the system).  We employed the ZTS method using our sampled trajectories as input.
Specifically, we linearly interpolated between every other time-slice in the sampled diffusive trajectory
and performed the ZTS minimization using a path composed of $n/2$ images of the system.  
The ZTS calculations were performed in the absence of the cluster constraining potential, $V_c$.  Convergence was determined to be reached when the "minmax" image, namely the highest energy image along the quenched path, exhibited an RMS force of less than $0.001$ per atom.  
The potential energy profile for a typical transition path that has been quenched in this fashion, and which illustrates the ruggedness of LJ$_{38}$ potential energy suface, is shown in Figure \ref{afterquench}.

\begin{figure}[!tbp] 
\includegraphics[angle=270,width=8.5cm]{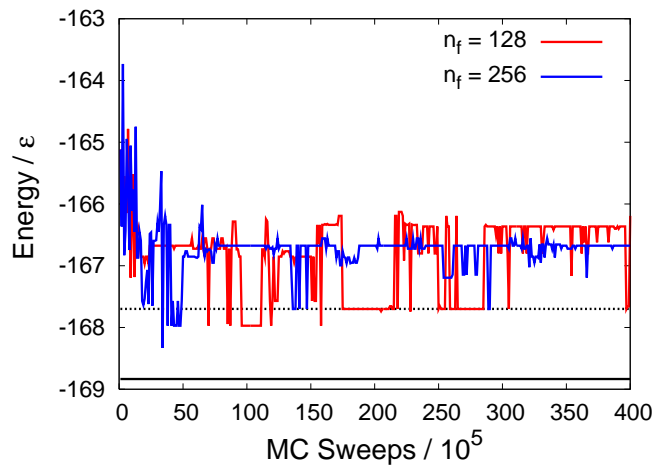} 
\caption[sqr]
{\label{minmax}
Minmax values for LJ$_{38}$ interconversion trajectories sampled at $T=0.1$ using the S\&S algorithm with fragment lengths $n_f$=128 (red) and $n_f=256$ (blue).  Trajectories are output every $10^5$ Monte Carlo sweeps.  The solid and dashed black lines indicate the minmax values for the lowest energy path and most probable path at T=0.02$\epsilon$, respectively, according to a DPS calculation by Wales.\cite{WalPER}  The DPS most probable path is determined by a harmonic rate theory.\cite{Wal02}
}
\end{figure}

In Fig. \ref{minmax}, we present the time series of minmax values obtained from the two S\&S simulations at $T=0.1$ reported here.  The red and blue lines indicate the results for the $n_f=128$ and $n_f=256$ simulations, respectively.  Trajectories recorded during the equilibration stage of the simulation are included.  The figure also includes two results from a Discrete Path Sampling (DPS) calculation\cite{Wal02} provided by Wales.\cite{WalPER}  The solid black line indicates the lowest minmax value found to connect the permutational isomers, an energy of $-168.8346$.  The dashed line indicates the minmax value of $-167.7005$ for the most probable path at the lower temperature of $T=0.02$, according to the harmonic rate theory employed in the DPS calculation.

It is clear that the S\&S algorithm at least partially succeeds in sampling the distribution of paths.  It would not at all have been surprising if the sampler had simply gotten trapped in the region of high-energy pathways around the initial straight-line trajectory.  Instead, within the first $25$ output trajectories, it finds its way to lower minmax values and proceeds to explore many different transition pathways.  
It is especially encouraging that, upon finding a particular low-energy pathway, the sampler does not become endlessly trapped.  In many instances, we see that a low minmax value is found, but then abandoned soon after in favor of higher values.
This observation also suggests that finite temperature, and perhaps the finite trajectory time $t$, play a role in biasing the ensemble of trajectories away from the global minimum energy path.

The key role of temperature, and thus entropy, in shaping the transition path ensemble is corroborated by the DPS calculations.  Even at the exceedingly low temperature of $T=0.02$, the lowest energy transition path (solid line) is not predicted to be the most favorable (dashed line).  It is therefore not surprising that our S\&S simulations, which sample trajectory space at the phase change temperature of $T=0.1$, prefer transition paths that are still higher in minmax energy.  Both S\&S  simulations do locate the pathway indicated by the dashed line in the figure, but they quickly move on in favor of more energetic paths.  These observations highlight the importance of methods that actually sample trajectory space.

Despite these encouraging observations, we can not claim that the S\&S algorithm achieves fully ergodic sampling in either simulation.  
In both S\&S simulations reported in Fig. \ref{minmax}, we see extended periods in which the sampler, although not necessarily locally trapped, appears to be confined within a "tube" of accessible trajectories.  For example, beyond the first 100 output trajectories in the $n_f=256$ simulation,  we see that the system consistently returns to a minmax value of $-166.675$.  A similar recurrence in the $n_f=128$ simulation is seen at the minmax value of $-166.364$.
These apparent transition tubes are analogous to the basins of stability found in configuration space, and the sampler's task of equilibrating them is difficult for the same reasons.
Further characterization of these transition tubes, perhaps using the finite temperature string method in coarse variables,\cite{Mar06} is an interesting possibility for future work.

Sweep for sweep, the $n_f=256$ simulation shown in the blue should do a better job of sampling than the $n_f=128$ simulation shown in the red. The former employs twice the fragment length and thus attempts all of the same Monte Carlo moves, plus those associated with its most coarse LC layer.  That the two simulations do not sample the same transition tube is an indication that the $n_f=128$, at least, is not ergodic, and most likely neither is.   
However, given the extreme ruggedness of the LJ$_{38}$ potential energy surface and the complexity of its low-temperature interconversion process, it is very encouraging to find that reasonable transition paths are obtained using the S\&S algorithm and that the local sampling of path space, at least, is achieved.

\section{Summary and Conclusions}

In this work, we have described several developments that aide the efficient transition path sampling of double-ended diffusive trajectories.  
In particular, we have introduced the symmetric Onsager-Machlup formula, which can be utilized as the basis for deriving high-order symmetric integrators that utilize only the forcefield, and we have shown that a direct time discretization on a uniform grid generates an approximation that is already accurate to second order.
Furthermore, we have shown that the sliding and sampling algorithm can naturally leverage massively parallel computer architectures to simulate double-ended trajectories on long timescales.

 Using a two-dimensional example, we demonstrate the computational and wall-clock efficiency of the S\&S algorithm.  We find that, because of the layered (multi-scale) structure of the LC representation and the fact that the sliding moves eliminate the intrinsic asymmetries of this representation,
the computational efficiency of the S\&S algorithm is relatively insensitive to the length of fragments employed.
This observation greatly simplifies the implementation of the algorithm - the fragment lengths can be primarily chosen according to the number of available nodes and the desire to minimize the wall-clock time of simulation.

We also employ the S\&S algorithm to study the low-temperature structural interconversion of the LJ$_{38}$ cluster.  We find that, for the simulation times employed here, the S\&S algorithm did not ergodically sample the full trajectory space - a result that is not at all surprising given the difficulty of equilibrating the low-temperature thermodynamics of this system with the direct Metropolis algorithm.
However, we are encouraged by finding that local sampling of path space is achieved with the S\&S algorithm and that the sampler was able to find a large number of low-energy transition pathways.

 Biological systems at room temperature do not exhibit the same free energy ruggedness that is found in the LJ$_{38}$ cluster.
 But they can still have metastable intermediates, entropic bottlenecks, and exceedingly long timescales, which frustrate open-ended trajectory sampling algorithms.
 The success of the direct Metropolis Monte Carlo algorithm in sampling the configuration space of such systems is a good indication that the S\&S algorithm will be similarly successful in sampling their trajectory space.
 It is in this regime of biologically relevant processes that we expect the S\&S algorithm to find its most immediately applicability.
 
\begin{acknowledgments} 
This work was supported in part by the National Science Foundation Grant No. CHE-0345280 and by the Director, Office of Science, Office of Basic Energy Sciences, Chemical Sciences, Geosciences, and Biosciences Division, U.S. Department of Energy under Contract No. DE-AC02-05CH11231.  The authors thank D. Chandler, J. D. Doll, E. Vanden-Eijnden, and W. H. Miller for many helpful discussions.  Special thanks is given to D. J. Wales for providing the results of his DPS calculations.
\end{acknowledgments}

\end{document}